# Snow particle analyzer for simultaneous measurements of snow density and morphology

Jiaqi Li,[a,b] Michele Guala,[a,c] Jiarong Hong.[a,b]

[a] *Saint Anthony Falls Laboratory, University of Minnesota, Minneapolis, MN*

[b] *Department of Mechanical Engineering, University of Minnesota, Minneapolis, MN*

[c] *Department of Civil, Environmental, and Geo- Engineering, University of Minnesota, Minneapolis, MN*

*Corresponding author*: Jiarong Hong, jhong@umn.edu

ABSTRACT

The detailed characterization of snow particles is critical for understanding the snow settling behavior and modeling the ground snow accumulation for various applications such as prevention of avalanches and snowmelt-caused floods, etc. In this study, we present a snow particle analyzer for simultaneous measurements of various properties of fresh falling snow, including their concentration, size, shape, type, and density. The analyzer consists of a digital inline holography module for imaging falling snow particles in a sample volume of 88 cm$^3$ and a high-precision scale to measure the weight of the same particles in a synchronized fashion. The holographic images are processed in real-time using a machine learning model and post-processing to determine snow particle concentration, size, shape, and type. Such information is used to obtain the estimated volume, which is subsequently correlated with the weight of snow particles to estimate their density. The performance of the analyzer is assessed using monodispersed spherical glass beads and irregular salt crystals with known density, which shows <5% density measurement errors. In addition, the analyzer was tested in a number of field deployments under different snow and wind conditions. The system is able to achieve measurements of various snow properties at single particle resolution and statistical robustness. The analyzer was also deployed for four hours of operation during a snow event with changing snow and wind conditions, demonstrating its ability for long-term and real-time monitoring of the time-varying snow properties in the field.

SIGNIFICANCE STATEMENT

Our study introduces a powerful, low-cost, compact instrument to simultaneously measure the morphological properties and density of fresh falling snow in situ. Such an instrument can be applied to various snow conditions and is capable of resolving the detailed morphological features and density of individual particles. In addition, the data processing software of the instrument based on machine learning allows automated, high-throughput, and robust analysis in real time. Our system provides critical information for the modeling of snow settling and the linkage between the precipitation forecast and ground snow accumulation. It can also be extended for measurements of other particles in industries and natural processes, such as mineral dust, embers, sediments, volcanic ashes, and pollens.

# 1. Introduction

Snow density and morphology have been observed to exhibit significant variability under different meteorological conditions (Magono & Lee 1966, Heymsfield 1972, Li, Cheng et al. 2021). It is therefore critical to measure both of them simultaneously near the ground to better forecast the snow accumulation rate and to estimate the snow water equivalent (SWE) for hydrology and water conservation studies, for prediction and prevention of snow hazards (e.g., avalanches and snow drift over transportation infrastructures), and more broadly for studying the evolution of the snowpack and the long term impact of snow cover on floods and climate. Specifically, the rate of snow accumulation on the ground and its spatial variability largely depend on the terminal velocity of falling snow particles, which is also observed to vary significantly in space and time over both synoptic and micro-meteorological scales (Heymsfield 1972, Garrett & Yuter 2014, Li, Cheng et al. 2021). Density and morphology of fresh falling snow are also important parameters to determine the properties of layered snowpack such as the SWE (Jonas, Marty & Magnusson 2009, Sturm et al. 2010), forcing on the ground, as well as conductive and radiative thermal properties (Sturm et al. 1997, Haussener et al. 2012). These properties are essential for predicting and preventing snow hazards such as avalanches (Steinkogler, Sovilla & Lehning 2014) and transportation accidents (Ogura et al. 2002). Furthermore, the thermal properties (i.e., conductivity, reflectance, and transmittance of snowpack) are expected to impact the local climate (Cohen & Rind 1991) and snowmelt-caused floods (Marks et al. 1998).

The terminal velocity of snow is a required modeling input to link the weather precipitation forecast with predicted accumulation on the ground. It is a very difficult quantity to be determined precisely due to large uncertainties in hydrometeor size, density, and drag properties, in addition to particle turbulence interaction mechanisms. In Nemes et al. (2017), Li, Cheng et al. (2021), and Li, Jiaqi et al. (2021), inertial properties of snow particles have been indirectly estimated using the distribution of the snow particle velocity acceleration. Since the work of Bec et al. (2006), it was observed that fluid parcels, i.e., passive tracers, respond to fluid acceleration much more dynamically than heavier particles and thus exhibit thicker tails in the acceleration PDF. The particle inertial properties are quantified by the response time $\tau_p$, i.e., by the time it takes to respond to a variation in fluid velocity, which depends on the particle density, size, and drag coefficient. In the simpler case of a spherical shape at very low Reynolds number and thus Stokes drag has an analytical formulation (Clift, Grace & Weber 2005). More complex particles, settling at high Reynolds number, such as snow particles, exhibit a non-linear drag coefficient and their settling velocity $W_s = \tau_p g$ remains unknown ($g$ is the gravitational acceleration). For snow, even in the absence of turbulence, disentangling the density and size effects on $W_s$ from the drag coefficient is challenging because of the variability in snow morphology and drag area (see for instance Tagliavini et al. 2021 & 2022). Therefore, snow settling prediction must rely on two possible modeling avenues: first, an empirical, indirect estimate of $\tau_p$ from particle acceleration (e.g., Nemes et al. 2017), assuming turbulence effect can be included through the dimensionless Stokes number $St = \tau_p/\tau_\eta$ (where $\tau_\eta$ is the Kolmogorov time scale); second, a direct estimate of $\tau_p$ based on measurements of snow density and morphology and assumptions on the drag coefficient. In order to explore the latter method, the specific shape and weight of snow particles must be provided to estimate the correct drag area and particle equivalent diameter, thus enabling the distinction among spherical particles (graupels and small particles), cylinders (needles), porous disks (plates and dendritic crystals), and porous aggregates of complex shape. Hence, in addition to image-based methods for estimating snow settling speed in the field (Nemes et al. 2017, Li, Cheng et al. 2021, Li, Jiaqi et al. 2021), it is important to deploy instrumentation designed to estimate snow particle size, shape, and density.

Currently, most snow analysis tools are designed to measure either the snow density or morphology, not both at the same time. Commonly used measurement techniques in the field are usually designed for measuring the density of accumulated snowpack, including snow wedge cutters (Conger & McClung 2009, Proksch et al. 2016), snow tubes (Hribik et al. 2012, Zhang, Su & Wang 2017), snow fork and micro-CT scan (Elder et al. 2019). Snow wedge cutters are stainless steel boxes with different shapes. They are used to enclose the snowpack filling the box and measure the density ($\rho$) based on the gravimetric measuring method ($\rho = M/V$, where $M$ is the total mass of collected snow, and $V$ is the volume). Snow tubes use the same measuring method; however, they are standard tools used in the US (Federal Sampler), Russia and China (Model VS-43). Studies have found no significant difference among the resulting snowpack densities depending on the cutter or tube shapes (Proksch et al. 2016, Zhang, Su & Wang 2017). Snow fork, on the other hand, utilizes microwave resonance from 500 to 900 MHz to accurately measure the complex dielectric constant of snowpack around the measurement tip (Sihvola & Tiuri 1986, Elder et al. 2019). With empirical equations, both the density and the wetness (liquid water content) of the snowpack can be obtained. The micro-CT (computer tomography) scan can reconstruct the 3D representation of the measured snowpack, thus obtaining the grain size of individual snow particles and the overall snowpack density (Proksch, Löwe & Schneebeli 2015, Elder et al. 2019).

For the measurement of snow morphology, image-based techniques have been designed and employed, such as the hydrometeor velocity and shape detector (HVSD) and the multi-angle snowflake cameras (MASC). Barthazy et al. (2004) developed the HVSD for the measurement of snow shape and fall speed using two line-scan cameras. However, it is designed for hydrometers with a diameter > 1 mm, and the limited resolution (0.15 mm) leads to large uncertainties in size measurement for sub-millimeter particles. The MASC system utilizes three cameras, with the resolution ranging from 9 to 37 μm, viewing from different angles to obtain the 3D features of snowflakes (Garrett et al. 2012). It is also equipped with near-infrared motion detectors for snow fall speed measurements. This system has become a commercial tool for quantification of snow particle size, the 3D features, and settling velocity. It has been applied for characterizing the size and shape of snow aggregates (Jiang et al. 2019) and analysis of the Arctic precipitation (Fitch & Garrett 2022). In general, such a multi-camera system is susceptible to the small change in the camera location, which may require periodic re-calibration, limiting its implementation for long-term particle analysis in the harsh environment (e.g., snowfall events) in the field.

To simultaneously measure the density and morphology of fresh falling snow, researchers have employed aircraft equipped with imaging probes for size and shape measurements and counterflow virtual impactor (CVI) for snow water content measurements (Heymsfield et al. 2004). The imaging probes are line-scan cameras capturing the 2D projection of the snow particles, and the CVI measures the vapor content of the evaporated snow particles captured by it. The imaging probes can measure snow particle sizes as small as 33 μm, and the CVI can measure the weight of water droplets or ice crystals larger than 8 μm. In their method, the snow particle size-dependent concentration is assumed to be a single gamma distribution, and the particles are approximated as simple spheres enclosing them, which can potentially lead to large uncertainties. Moreover, the equipment for measuring snow particle size (imaging probe) and mass (CVI) work asynchronously (i.e., they do not necessarily measure the same snow particles). More recently, Singh et al. (2021) developed a Differential Emissivity Imaging Disdrometer (DEID) consisting of a thermal camera and a metal hotplate for measuring the mass, type, and density of hydrometeors. They utilized the large thermal emissivity difference between the hydrometeors (water) and the hotplate (metal) to measure the spatial dimension of the hydrometeors and converted the heat loss from the hotplate to evaporate them to mass. They tested the system in the lab and field (Rees et al. 2021), demonstrating that this disdrometer is insensitive to varying environmental conditions. However, the spherical approximation of the hydrometeor volume may lead to significant uncertainties for density measurements.

To address the limitations of the abovementioned snow analysis approaches, in this paper, we design and fabricate a snow particle analyzer by integrating digital inline holography (DIH) with a high-precision scale for simultaneous measurements of the density and morphology of falling snow particles near the ground. DIH has emerged as a compact tool for measurements of the 3D flow field and particle transport since the beginning of this century (Katz & Sheng 2010). It employs a coherent light source to illuminate a sample volume and capture the fringe patterns generated by the interference between the scattered signals from the sample and the unscattered portion of the illumination light source (referred to as holograms hereafter). The holograms contain information on the size, shape, and 3D position (longitudinal and lateral) of the particles in the sample volume. Such information can be extracted through conventional holographic reconstruction algorithms (e.g., Fraunhofer, Fresnel-Kirchhoff, or Rayleigh-Sommerfeld) or machine learning based on algorithms introduced recently by Shao et al. (2020a & 2020b). DIH has been widely applied for measuring mineral dust particles (Gaudfrin et al. 2020), cloud particles (e.g., cloud droplets, ice crystals; Fugal et al. 2004, Larsen et al. 2018), microorganisms (e.g., plankton Guo et al. 2021, and microcystis aeruginosa You et al. 2020), and drosophila (Kumar et al. 2016). Field measurements of snow particles with a large range of size and shape using DIH have also been successfully conducted with a simple and low-cost setup (Nemes et al. 2017, Li, Cheng et al. 2021, Li, Jiaqi et al. 2021). However, the previous system only measures the size of snow particles with no quantification of snow particle shapes and types as well as their masses for density calculation. In addition, it operates at a low sampling rate, and its hologram processing and snow particle extraction algorithm require largely manual processing, which significantly limits its ability to generate large datasets of snow particle holograms and for long hours of operation in the field. Our current work will be an extension of these past efforts by introducing a new generation of DIH-based snow particle analyzer, which employs highly efficient automated data processing and incorporates high-precision weight measurements for snow density estimate.

The design of the proposed snow particle analyzer, including the hardware and data processing software, is introduced in Section 2. In Section 3.1, we assess the performance of the snow particle analyzer using glass beads and salt crystals. The assessment using actual snow particles is introduced in Section 3.2. We then demonstrate the ability of our designed snow particle analyzer during actual snow events in Section 4, with the summary and discussion following in Section 5.

## 2. System Design

*a. Hardware design*

As shown in figure 1, the snow particle analyzer is comprised of digital inline holography (DIH) system for imaging, a high-precision scale for weight measurement, a laptop for synchronizing and controlling the data acquisition and processing of the DIH system and the scale, and additional components including power supply, shielding and support. Except for the laptop, all the other components are enclosed in a wooden box with the dimension of 24 cm x 24 cm x 37 cm.

The DIH system employs a 532 nm diode laser, a small concave lens (5-mm diameter, 5-mm focal length) as the beam expander, a collimating lens (bi-convex lens with 50.8-mm diameter and 75-mm focal length), a condenser lens (aspherical condenser lens with 50.8-mm diameter and 32-mm focal length), an imaging lens (Fujinon CF25HA-1 25 mm lens) that provides 0.5x magnification, and a CMOS camera (Teledyne FLIR, Blackfly S USB3 mono, model: BFS-U3-123S6M-C). During image capturing, the camera sensor uses an active area of 2048 x 1500 pixel$^2$ after decimation by a factor of 2 in both width and height, operating at the frame rate of 50 frames per second (FPS). Correspondingly, the spatial and temporal resolutions of our DIH system are 14.3 μm/pixel and 0.02 s, respectively, with a sample volume of 2.93 x 2.15 x 14 cm$^3$.

The high-precision scale (Vibra HT Series from Intelligent Weighing Technology) with a 0.1 mg resolution, a 200 g capacity, and a 10 Hz sampling rate is installed under the DIH system. A transparent snow sample collector is placed on the scale with the opening leveled with the upper side of the sample volume to make sure all snow particles captured by the imaging system fall into the collector to be weighed. The whole integrated system is enclosed in a wooden box for insulation and waterproofing. The top opening is adjustable with a maximum size of 4.5 cm x 14 cm. Two shields are 3D-printed to protect the extended part of the DIH system. The power supplies for the laser and the scale are secured inside the box. A tripod adapter at the bottom enables easy deployment in the field. Both the camera and the high-precision scale are connected to a laptop through USB3 cables, which uses a custom-designed data acquisition software based on Python to control the synchronization and data acquisition of the camera and the scale.

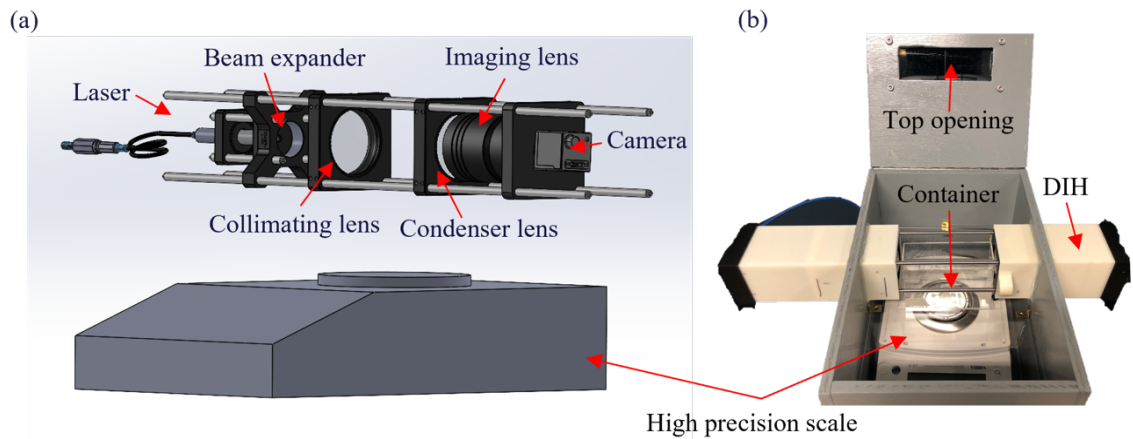

Fig. 1. (a) A schematic and (b) a photo of the snow analyzer constructed by integrating a DIH system with a high precision scale.

*b. Data processing method*

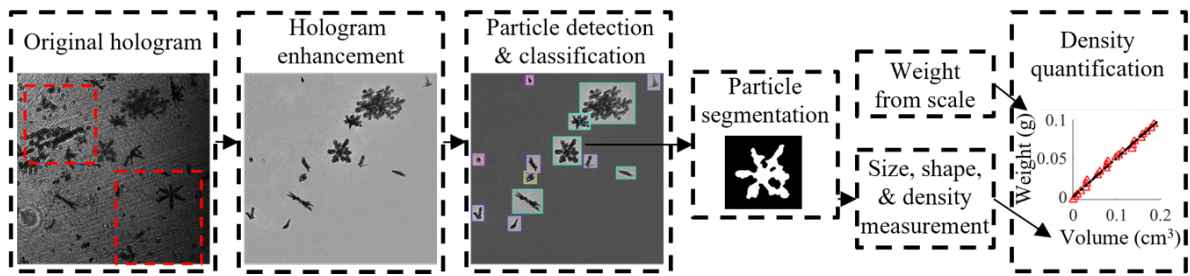

Fig. 2. Illustration of data processing procedure used to extract snow particle properties from DIH images, including hologram enhancement, particle detection and classification, particle segmentation, size and shape measurement, and density quantification. The red dashed boxes mark the particles sticking to the collector wall (static background) that are removed after hologram enhancement. The inset plot in the density quantification step corresponds to the linear-fitting-based average density measurement method.

The data processing is divided into five steps, including hologram enhancement, snow particle detection and classification, snow particle segmentation, size and shape measurement, and density quantification, as shown in figure 2. The collected holograms are first enhanced to remove the static background and other noises. Then, the enhanced holograms are fed into a trained machine learning model for snow particle detection. Classification is applied in parallel with the detection based on the snow particle morphology to better estimate the volume of snow particles from the dimensions of their segmented 2D projections. The volume and snow particle type are finally correlated with the weight measurement from the high-precision scale for density quantification.

A machine-learning-based approach is employed for the detection and classification of snow particles. We apply the YOLOv5 (You Only Look Once) architecture for these tasks (Jocher et al. 2020), which is an improved version since the first release of YOLO (Redmon et al. 2016) based on the Pytorch framework. YOLO is a unified, general-purpose model for object detection (including simultaneous classification). The architecture consists of a backbone for multi-stage feature extraction, a neck for combining these multi-stage features, and a head for predicting the bounding boxes and class of the objects. Since it is a unified model and a one-stage detector, YOLO models usually have higher processing speed as compared to other object detection models while maintaining high precision, allowing real-time image processing for a broad range of applications (Diwan et al. 2022). Specifically, the YOLOv5 model trained on the COCO dataset (Lin et al. 2014) reaches more than 50% average precision (AP) at more than 100 frames per second processing speed using Nvidia Tesla V100 GPU. In this study, we customize the feature extraction layers by employing fewer layers, and the feature connecting methods in the neck to accommodate the characteristics of holograms and for even faster processing speed. As shown in table 1, the snow particles are classified into six categories with assigned codes based on Magono & Lee (1966): aggregate/irregular (I), dendrite (P2), graupel/rime (R), plate (P), needle/column (N/C), small particles/germ (G). Sample holograms for each type of snow particle are presented in figure 3. A total of 2500 snow particles (each in a 256 x 256 pixel$^2$ image) are handpicked and manually classified as the training dataset. Each 16 snow particles are randomly selected to form a 1024 x 1024 pixel$^2$ combined image, and a total of 520 images are generated for training. The images are annotated using the software Roboflow (Roboflow Inc., https://roboflow.com/) with a bounding box for each snow particle and its type. Data augmentation is applied to the raw images to enrich the dataset by rotating the images by 90 degrees, changing the exposure between ±30%, blurring up to 7 pixels, and adding salt and pepper noise to up to 5% of the pixels. We train the model for 800 epochs, and the snow particle detection rate reaches 99%, with an average classification accuracy of 95%.

The detected snow particles are reconstructed to the focal plane, and threshold-based segmentation is applied to binarize the snow holograms. We conduct region-based measurements to obtain the size and shape of the detected snow particles. Our system is designed to operate at a frame rate of 50 FPS, which allows us to capture snow particles in two or more consecutive frames considering the maximum fall speed of snow particles observed in our experiments. Accordingly, our data processing method incorporates a particle filtering step that uses the size, shape, and average displacement of snow particles to remove those captured more than once. The volume of each snow particle is then estimated based on equations in table 1. For aggregates, due to the complex shape, we first segment out the individual snow crystals (referred to as monomers in Dunnavan et al. 2019) that form the aggregates using the *watershed* function in MATLAB and treat each monomer as a spheroid with the diameter as the equivalent diameter ($d_{eq,i}$) of the 2D projection of the segment. The volume is thus estimated as the sum of the volume of the monomers. For graupels and small particles, we estimate the volume as a whole spheroid considering the simpler geometry following the convention in literature (Heymsfield et al. 2004, Singh et al. 2021). While for dendrites, plates, and needles, as one dimension of these particles is much smaller than the other dimensions (for needles, the dimension of the cross-section is smaller than the length; and for plates and dendrites, their thickness is much smaller), their volumes are estimated based on the cylindrical shape assumption. The major axis ($d_{maj}$) is used as the diameter for dendrites and plates, and their thickness ($T$) is obtained from empirical equations (Auer & Veal 1970). Needles, on the other hand, have the minor axis ($d_{min}$) as their diameter and the major axis ($d_{maj}$) as their length. The volume of individual snow particle from each frame of hologram, estimated accounting for specific snow morphologies, is then correlated with the weight signal from the high precision scale. As the volume measurement from holograms and the weight measurement are synchronized, we then divide between increments of weight and volume during selected sampling periods to obtain the density. The duration of the sampling period is selected considering the particle fall speed, scale response time, and wind-induced fluctuations. The minimal duration for the snow particle density measurement depends on the wind speed: one second under no wind allowing a single particle to be imaged and weighed, and 30 seconds under 2 m/s wind speed to average out fluctuations in the weight signal. We thus obtained the average density of all the snow particles captured during the sampling period. Thus, there are chances that only one snow particle is captured during the sampling period. This probability is higher under no wind, but much lower for higher wind speed and snow concentration. Moreover, the average density of multiple snow particles collected through an extended period can also be estimated as the slope of the linear fitting between the time-synchronized weight and volume signals (see example in the inset plot in figure 2).

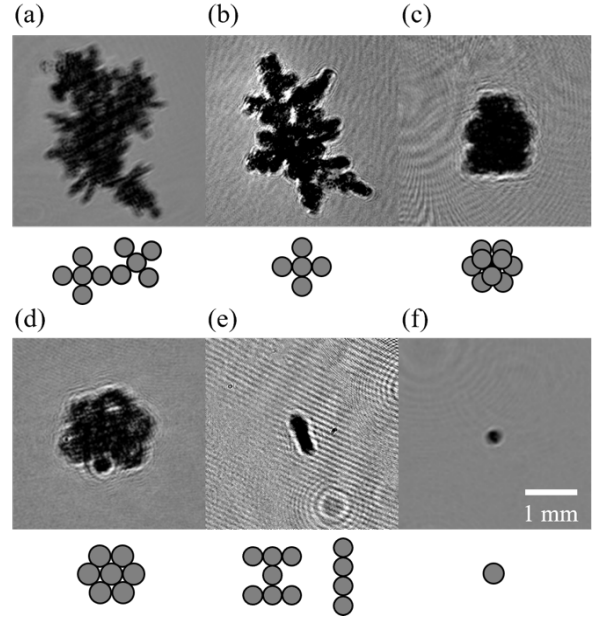

Fig. 3. Sample holograms and corresponding illustrations for each type of snow particles: (a) aggregate (I); (b) dendrite (P2); (c) graupel (R); (d) plate (P1); (e) needle (N/C); (f) small particle (G).

| Snow type | Code | Thickness (μm) | Volume |
|---|---|---|---|
| Aggregate | I | NA | $\sum_{i}^{N} \frac{1}{6}\pi d_{eq,i}^{3}$ |
| Dendrite | P2 | $T = 2.801 d_{maj}^{0.377}$ | $\frac{1}{4}\pi d_{maj}^{2} T$ |
| Graupel | R | NA | $\frac{1}{6}\pi d_{eq}^{3}$ |
| Plate | P1 | $T = 2.506 d_{maj}^{0.398}$ | $\frac{1}{4}\pi d_{maj}^{2} T$ |
| Needle | N/C | NA | $\frac{1}{4}\pi d_{min}^{2} d_{maj}$ |
| Small particle | G | NA | $\frac{1}{6}\pi d_{eq}^{3}$ |

Table 1. Classification and corresponding volume estimation of snow particles. The codes are assigned based on Magono & Lee (1966).

## 3. System assessment

*a. Assessment with glass beads and salt particles*

We first assess the performance of our snow particle analyzer in the laboratory using glass beads (spherical shape) and salt crystal particles (irregular shape) to quantify the uncertainties related to the size and density measurement. The calibration of the system is conducted by measuring the size and density of individual glass beads and compare with the reference density of $2.5 \times 10^3$ kg/m³ (Weast 1981). The glass beads show a relatively uniform size of $d = 1.36 \pm 0.06$ mm (figure 4a). Figure 4b shows the step increment of the volume and weight signals of each glass bead. We tested 20 individual glass beads, and the resulting average density is $2.52 \pm 0.13 \times 10^3$ kg/m³, showing 5% uncertainty in the density measurement. Measurement uncertainties can be attributed to the volume estimation (non-uniform shape) and fluctuations in the weight signal.

To assess the accuracy of our density estimation for particles appearing at different locations in the sample volume, we further test the system by dropping the glass beads at four different distances with respect to the camera imaging plane. Note that we also drop multiple glass beads at the same time to assess the ability of our system to handle high concentration cases. In this manner, the system is not able to accurately differentiate the weight signal of the individual particle, which is inevitable for occasions with high snow concentration. Thus, only the average density of the particles in the sample volume can be obtained. The average densities for each test are: $\rho_1 = 2.55 \times 10^3$ kg/m³, $\rho_2 = 2.44 \times 10^3$ kg/m³, $\rho_3 = 2.58 \times 10^3$ kg/m³, and $\rho_4 = 2.50 \times 10^3$ kg/m³, all within the 5% uncertainty range. Thus, we confirm the density measurement is insensitive to the relative distance of the particles to the camera.

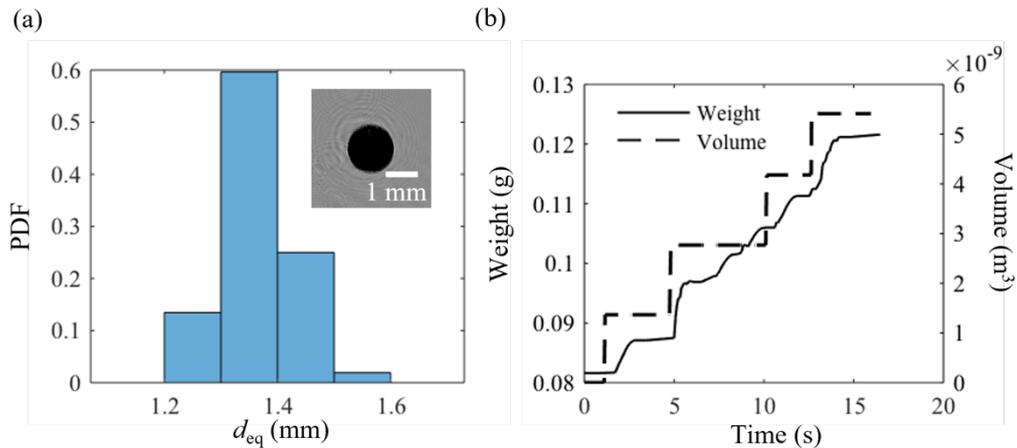

Fig. 4. (a) Size distribution of the glass beads, with a sample hologram; (b) step increment of the volume and weight signals for density measurement.

Subsequently, irregular salt particles are employed to further assess the performance of our snow particle analyzer for density, size, and shape measurements of particles with more complex geometries. Figure 5a shows sample enhanced holograms of salt particles. Due to the complex shape, we estimate the particle volume using $V = \frac{1}{6}\pi d_{eq}^3$, where $d_{eq}$ is the equivalent diameter of the 2D projection of imaged salt particles. In figure 5b, the average density is calculated as a function of the number of measured particles from the accumulated weight and volume using the equation $\bar{\rho}(n) = \sum_{i=1}^{n} m_i / \sum_{i=1}^{n} V_i$. Our density measurement initially fluctuates but gradually converges to a value of $2.07 \times 10^3$ kg/m³ when the number of salt particles used for density calculation increases above 12. The measured value is within 5% of the reference value of salty crystal density of $2.17 \times 10^3$ kg/m³ (Haynes, Lide & Bruno 2016). Note that the actual time to reach convergence can vary due to the particle concentration and overall mass being weighed.

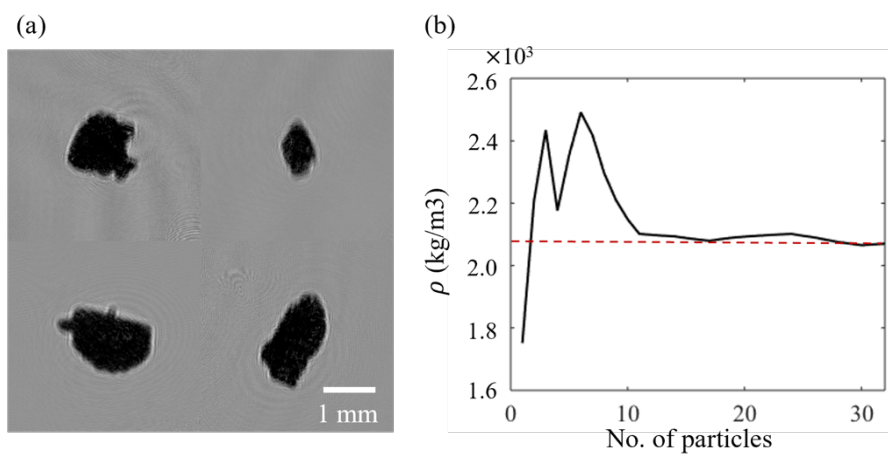

Fig. 5. (a) Sample holograms of the salt particles; (b) convergence of the average density with respect to the number of measured particles. The red dashed line shows the final average density.

*b. Assessment with snow particles of different types*

The snow particle analyzer is then assessed using snow particle hologram datasets collected during field deployments. Three datasets with different dominating types of snow particles, i.e., aggregates, graupels, and dendrites, are selected for the density calibration. These datasets are selected from multiple deployments of the snow particle analyzer at the Eolos field station in Rosemount, MN. The sample holograms from these datasets are shown in figure 6a, b, and c, respectively, illustrating the clear difference in morphology across the three types of snow particles. To account for the wind-induced fluctuations of the weight signal from the scale, we use linear fitting between the weight and volume time series to obtain the slope as the average density as demonstrated in figure 6d. The results indicate that the dendrites have the largest density while the aggregates are mostly porous and thus yield the smallest density. The percentages of the dominant types are 89%, 94%, and 85%, respectively for the aggregate-, graupel-, and dendrite-dominating dataset. To obtain a more accurate average density for each type of snow particle, we consider the impurities (other types of snow) within each dataset and solve a set of equations $\sum_i \rho_i V_i / V = \bar{\rho}$, where $\rho_i$ is the $i^{th}$ type of snow particles, $V_i$ is the corresponding volume, $V$ is the total volume increment throughout each dataset, and $\bar{\rho}$ is the average density obtained from the linear fitting. To reduce the number of unknowns, and to combine snow particles based on the volumetric estimates, dendrites and plates are grouped as one type, and graupels are grouped together with small particles (needles are neglected as there are less than 1% detected in all datasets).

The average density calculated for aggregates is $82 \pm 9$ kg/m$^3$, and it is comparable to the $20 - 150$ kg/m$^3$ density range from Ishizaka et al. (2016). We attribute this relatively lower density to the larger size and higher porosity of the aggregates in comparison to other examined types. We derive the average density for graupels to be $140 \pm 10$ kg/m$^3$. The measurement is in agreement with the density (~120 kg/m$^3$) measured by Ishizaka (1993) through microscopic imaging in the lab. These graupels have a relatively low average density, indicating porosity within the particles. Dendrites are estimated with a smaller volume (thin plate-like particle), with only air gaps between the icy branches. Thus, the dendrites have the largest density of $577 \pm 13$ kg/m$^3$, among the three types. Heymsfield (1972) summarized empirical equations for estimating the dendrite snow density as $\rho_{[g/cm^3]} = 0.588 d_{maj,[mm]}^{-0.377}$. With the size measured for the dendrites, we can obtain the average dendrite density in theory as 550 kg/m$^3$, comparable to the measured density. The density measurement uncertainties can be attributed to the errors involved in the 3D volume estimation, snow type classification, and the linear fitting method used to obtain the average density. Specifically, our 3D volume estimation shows 5% uncertainty based on the laboratory calibration using irregular salt particles and up to 4.9% uncertainty in volume considering the misclassification of snow type.

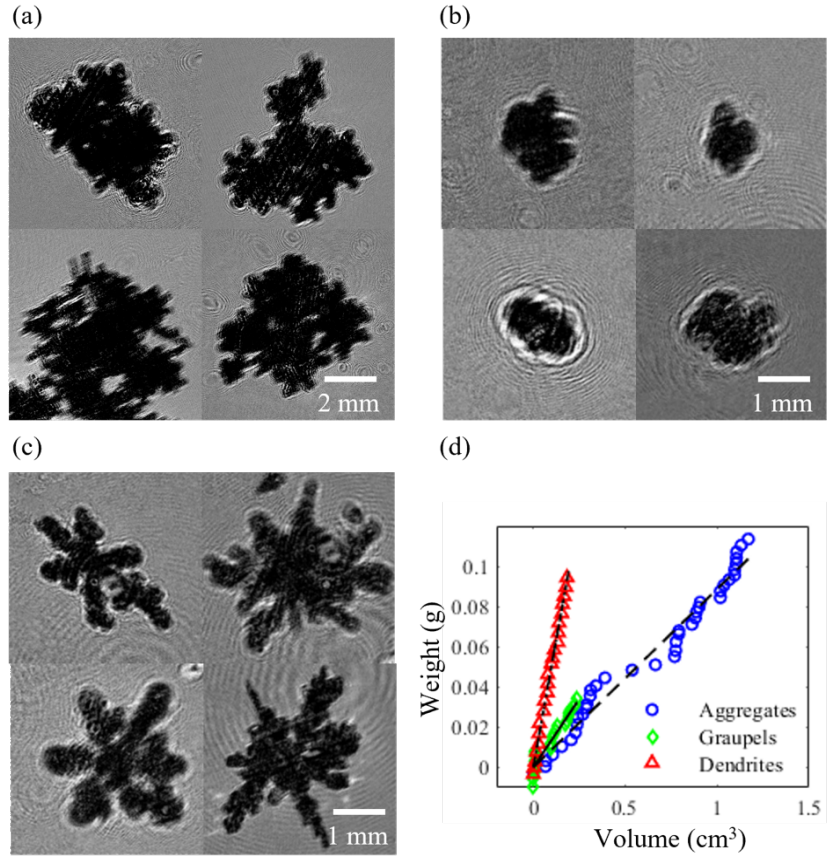

Fig. 6. Sample holograms of (a) aggregates, (b) graupels, and (c) dendrites; (d) linear fitting of the weight and volume measurements for the average density estimate.

*c. Assessment with individual snow particles*

Furthermore, the snow analyzer is assessed on its ability to quantify the density of individual snow particles. Individual particles are filtered out from the collected snow holograms by the following criteria: (1) only one particle is present in the current frame; (2) no other particles appear during the period $\pm 0.5$ s from the current frame considering the response time of the scale. We then obtain the weight from the scale time series, where an identifiable increment of the signal is found within 0.5 s after the snow particle is captured by the DIH system. In figure 7a, we show the time series of the weight signal associated with the corresponding imaged snow particle. Classification is also applied to the individual snow particles for better volume estimation. Our measurement results are shown in figure 7b and c. We collect particles with a large size range (from 0.5 mm to 6.5 mm), and individual needles are not observed after filtering due to the rare occurrence. The measured densities for each type are in agreement with the bulk measurement. Overall, we confirm the negative correlation between the snow particle density and size as reported in previous research (Heymsfield 1972,

Ishizaka 1993). Specifically, as shown in figure 7d, for dendrites (P2), the density empirical equation $\rho_{[g/cm^3]} = 0.588 d_{maj,[mm]}^{-0.377}$ from (Heymsfield 1972) with 34% average deviation defined as: $\frac{1}{N}\sum_{i=1}^{N} \frac{|\rho_{meas,i} - \rho_{calc}(d_{eq,i})|}{\rho_{calc}(d_{eq,i})}$, where $\rho_{meas,i}$ is the measured density of the $i$th snow particle, and $\rho_{calc}(d_{eq,i})$ is the calculated density using its equivalent diameter from the empirical equation. As no previously reported empirical equation is available for graupels and aggregates, we conduct power-law fitting of the size and density of these types of snow particles in the form of $\rho = a d_{eq}^{-b}$. We obtain a relation between the size and density of graupels and aggregates described by the equation $\rho_{[g/cm^3]} = 0.529 d_{eq,[mm]}^{-1.32}$ as illustrated in figure 7e, with a 55% average deviation of the data points from the proposed correlation. (explain the large deviation) Last but not least, the small particles (G) and plates (P1) have less porosity as compared to the aggregates, graupels, and dendrites. Thus, their densities tend to be close to the density of ice with little size dependence.

The above assessment proves that the snow classification and the associated density estimation are reasonably accurate at the single particle scale and not only statistically under snowfalls characterized by a dominant morphological type. Moreover, the individual snow density analysis can provide us with the necessary information to understand the impact of individual snow density and morphology on its settling behavior. With experimentally validated, size-specific, statistical correlations between snow particle size and density, we can estimate the density of specific snow particles using only imaged-based methods with no need for simultaneous weight measurements. Such correlations would allow us to extend the scenarios in which we can deploy our system for studying snow particle aerodynamic properties and improve snow settling predictions, for example, under high snow concentration, where only the average snow density of particles with different size and shape can be obtained, or under strong wind where weight measurements are challenging.

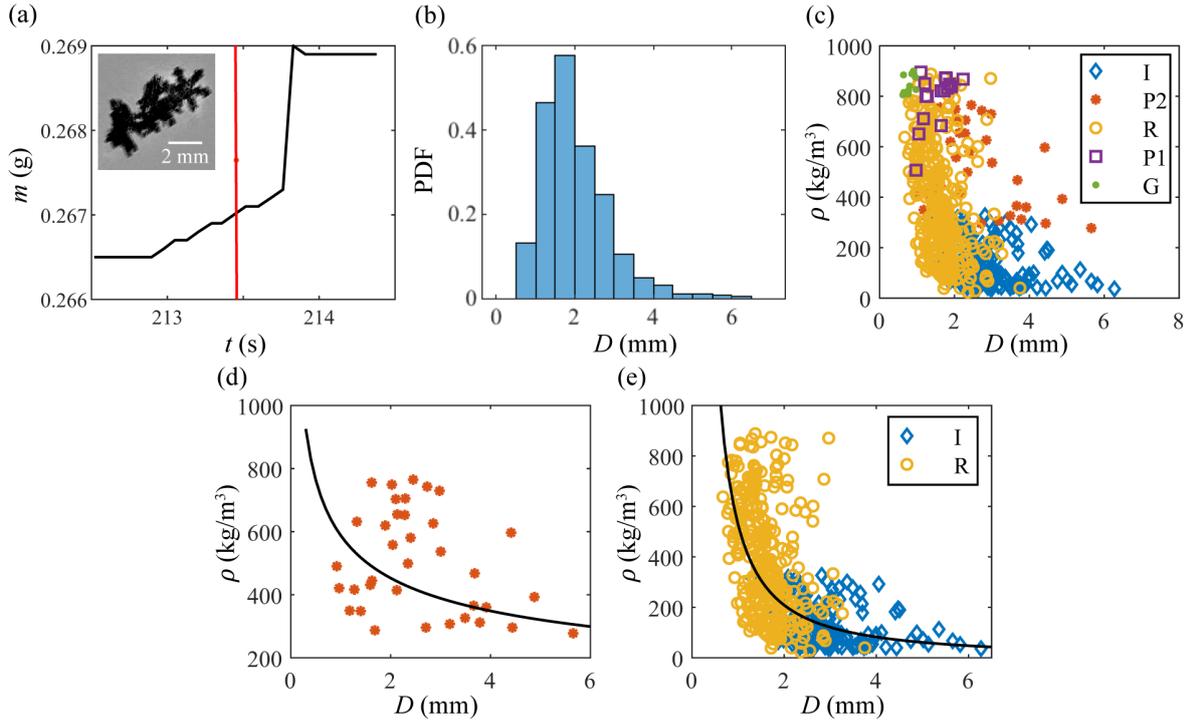

Fig. 7. (a) A sample time series of the weight signal from the high precision scale with the red line representing the exact time when a single snow particle (shown in the sample hologram) is captured by the DIH system; (b) probability density function (PDF) of the snow particle diameter ($D$ is defined as $d_{eq}$ for aggregates I, graupels R, and small particles G, and as $d_{maj}$ for dendrites P2 and plates P1); (c) scattered plot of the measured individual snow particle density and diameter ($D$); (d) scattered plot of measured density and diameter of the dendrites (red stars) and the empirical curve from Heymsfield (1972) (black line); (e) scattered plot of measured density and diameter of the aggregates (blue diamonds) and graupels (orange circles) and our proposed empirical curve (black line).

## 4. System demonstration

Finally, the performance of our snow analyzer is demonstrated through a multi-hour field deployment, during which the snow analyzer is shown to provide a detailed characterization of the variation of snow particle properties over time. The deployment was conducted on Jan. 22, 2022, between 18:00 and 23:00 local time, at the Eolos field station in Rosemount, MN. The top plot in figure 8 shows the streamwise wind speed and turbulent intensity variation over time. The blue dashed boxes indicate the three 9-minute sampling periods during which the snow particle analyzer was acquiring data. Under 50 FPS, a total of 24,000 holograms are captured for each sampling period. The wind speed increases at around 20:00 local time and stays at about 2 m/s after 20:30. Turbulent intensity first increases and drops at around 20:00, then grows again after 20:15 and holds at the highest level from 20:50 till the end. Although the wind changes during the deployment, temperature and relative humidity keep the same level at around $-13°C$ and 84%, respectively.

The snow type and size distribution across the three sampling periods change drastically, as shown by the six small plots in figure 8. Note that the snow particle diameter $D$ is defined as $d_{eq}$ for aggregates, graupels, and small particles, as $d_{maj}$ for dendrites and plates, and as length ($L$) for needles. Dendrites and plates dominate the snow particle types for sampling period 1, with a more uniform distribution in size across the three periods. The number of dendrites and plates decreases during sampling period 2, with a wider size distribution, indicating larger aggregates. Sampling period 3 has mostly graupels and few aggregates, with no particles with a diameter larger than 4 mm.

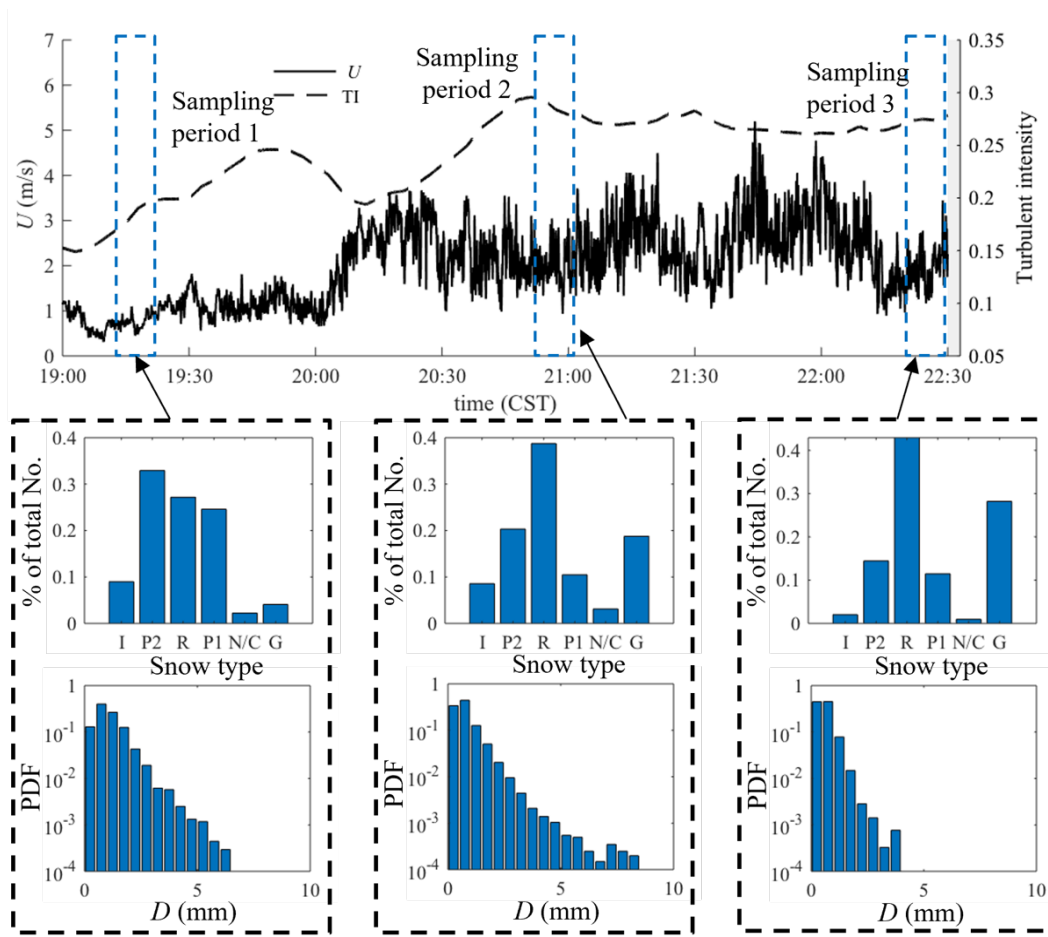

Fig. 8. Instantaneous wind speed ($U$) and turbulent intensity (TI) derived from moving-window average throughout the whole deployment event. Blue dashed boxes show three sampling periods with varying snow density and morphology. Inset plots show the distributions of the snow type and size ($D$ is defined as $d_{eq}$ for aggregates, graupels, and small particles, as $d_{maj}$ for dendrites and plates, and as length $L$ for needles) captured during the three sampling periods.

Specific time variations of the snow density and size are presented in figure 9. As discussed in section 2, we apply a 30-second moving averaging window to the time series of the particle size and density estimates for the three distinct sampling periods considering the wind-induced fluctuations of the weight signal. The average density decreases during sampling period 1 and climbs up during sampling period 2. During sampling period 3, the fluctuation magnitude of average density is shown to be much larger, due to more substantial variability in snow morphology, with larger density peaks representing short periods with smaller and denser particles. Note that empirical equations from Heymsfield (1972) suggest that the density of snow particles exhibits a negative correlation with their size, supporting our measurements here (comparing the temporal trends of $\langle \rho \rangle_{30s}$ and $\langle D \rangle_{30s}$, in figure 9a-9d, 9b-9e, while it is less clear in figure 9c-9f).

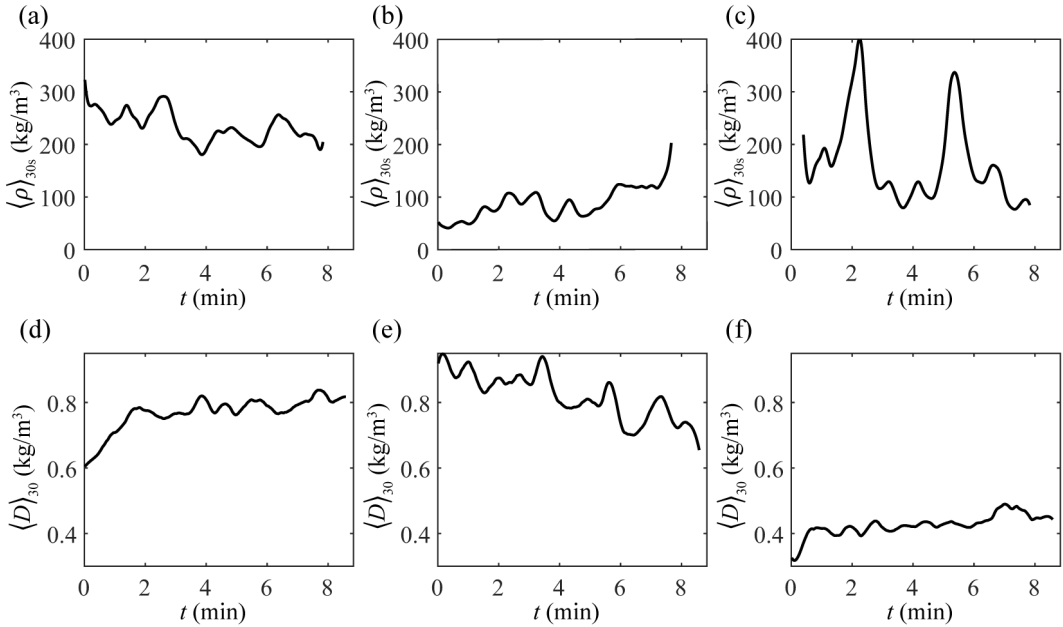

Fig. 9. The average density ($\langle \rho \rangle_{30s}$) calculated in a 30-second moving window over the sampling period 1 (a), 2 (b), and 3 (c); and the average snow particle diameter ($\langle D \rangle_{30s}$) calculated in the same fashion for sampling period 1 (d), 2 (e), and 3 (f).

The mean snow size ($\bar{D}$), density ($\bar{\rho}$), volume fraction ($\phi_V$), together with the mean wind speed ($\bar{U}$), root mean square of fluctuating velocity ($u_{rms}$), turbulent intensity (TI), relative humidity (RH), and temperature ($\bar{T}$) during the three sampling periods are listed in table 2 for comparison. Note that the volume fraction is obtained by the equation $\phi_V = V_{total}/(N_{img} V_S)$, where $V_{total}$ is the volume of all snow particles detected during the sampling period, $N_{img}$ is the number of images captured, and $V_S$ is the sample volume of the DIH system. The volume fraction estimated for the current snow event is comparable to those from Nemes et al. (2017), Li, Cheng et al. (2021), and Li, Jiaqi et al. (2021).

In comparison to the other two sampling periods, the snow particles from sampling period 1 have a relatively large size and the most significant density due to the presence of many dendrites and plates. The temperature of around −13.5°C also indicates primarily dendrite- and plate-like crystals form in the clouds (Magono & Lee 1966). As the snow volume fraction, wind speed, and turbulent intensity increase, snow particles have a higher chance of colliding and forming larger aggregates when they fall from the clouds to the ground during sampling period 2 (Fujiyoshi & Wakahama 1985, Dunnavan et al. 2019). The reduction in the number of dendrites and plates with large densities, as well as the growing percentage of large size particles, lead to the smaller snow density during sampling period 2 as compared to that during sampling period 1. During sampling period 3, near the end of the snow event, the concentration of snow in the atmosphere decreases as indicated by the drop of snow volume fraction, leading to a lower snow particle collision rate. Thus, the number of aggregates is at its lowest, and the snow particles are mostly small graupels generated by the deposition of water vapor on the snow crystals while falling from the clouds to the ground (Harimaya 1988). As the mean particle size decreases drastically from sampling period 2 to 3, along with the reduction in the number of aggregates characterized by smaller density, the mean snow particle density increases from 91 kg/m$^3$ to 200 kg/m$^3$.

| Sampling period | $\bar{D}$ (mm) | $\bar{\rho}$ (kg/m$^3$) | $\phi_V$ × 10$^{-7}$ | $\bar{U}$ (m/s) | $u_{rms}$ (m/s) | TI | RH | $\bar{T}$ |
|---|---|---|---|---|---|---|---|---|
| 1 | 0.77 ± 0.46 | 234 ± 24 | 15 | 0.86 | 0.17 | 0.19 | 84.4% | 13.1°C |
| 2 | 0.80 ± 0.57 | 91 ± 16 | 28 | 1.99 | 0.54 | 0.29 | 85.0% | 13.4°C |
| 3 | 0.43 ± 0.23 | 200 ± 38 | 22 | 1.98 | 0.57 | 0.28 | 84.6% | 13.0°C |

Table 2. The mean diameter and density of snow particles captured during the three sampling periods with the volume fraction of snow, mean wind speed, root mean square of the fluctuating velocity, turbulent intensity, and the maximum differences of relative humidity and temperature to their ensemble average values.

## 5. Conclusions and discussion

In this study, we present a snow particle analyzer for simultaneous measurements of various properties of fresh falling snow, including their concentration, size, shape, type, and density. The analyzer consists of a digital inline holography (DIH) system for imaging falling snow particles in a sample volume of 88 cm$^3$ and a high-precision scale to measure the weight of these same particles in a synchronized fashion. The holographic images are processed in real-time using a customized YOLOv5 machine learning model to determine snow particle concentration, size, shape, and type. Such information is used to classify the snow precipitation and correctly estimate the volume, which is subsequently correlated with the weight of snow particles measured by the scale to obtain their density. The performance of the analyzer is assessed using monodispersed spherical glass beads and irregular salt crystals with known density, which shows <5% density measurement errors despite the uncertainties involved in the particle volume estimate. In addition, the analyzer has been tested in a number of field deployments under different snow and wind conditions. The system is able to achieve measurements of various snow properties at single particle resolution and statistical robustness. Finally, the analyzer was also deployed for four hours of operation during a snow event with changing snow and wind conditions, demonstrating its ability for long-term and real-time monitoring of the time-varying snow properties in the field.

Our system is able to simultaneously measure the high-resolution morphology and weight, providing accurate density of fresh falling snow down to the single-particle resolution. In addition, the integration of machine learning-based data processing allows us to autonomously detect and classify the snow particles and measure the rich information such as the size, shape, and density of snow particles. With such measurements, we are able to establish statistical correlations among snow properties (i.e., size, type, and density) based on single particle data collected from these long-term deployments. Such information is critical for better modeling of the terminal velocity of falling snow, predicting snow accumulation on the ground, as well as the porosity, density, and thermal properties of the snowpack, and estimating the snow water equivalent for hydrology studies. Compared to the aircraft measurement system, our snow particle analyzer can measure the morphology and density of individual snow particles without assumptions about the size and weight distributions. Our system also provides more accurate volume estimation based on the classified snow type, with machine learning-based image processing enabling a much higher processing speed in comparison to the Differential

Emissivity Imaging Disdrometer. Moreover, the capabilities of our system can be extended to other particles in industries and natural processes where concentration, size, shape, type, and density are important, such as mineral dust, embers, sediments, volcanic ashes, and pollens, etc.

There are still several improvements that can be made to our design. First of all, the minimal duration of the sampling period to obtain stable density measurement is limited by factors such as the response time of the high-precision scale, the time lag due to particle falling, and wind-induced fluctuations. We can potentially use a high-precision load cell that has a faster response to the weight changes and improve the mounting of the collector on the scale to reduce this minimal duration. In addition, with high wind speed, the weight of snow particles may not clearly emerge from the large wind-induced temporal fluctuations. It would be problematic if the snow and wind conditions change within a short time scale over which the wind-induced weight fluctuations cannot be averaged out. Shields around the snow particle analyzer would be necessary to minimize such fluctuations in the weight signals. Second, snow terminal velocity is an important variable for predicting snow accumulation, which is not measured in our current system due to the confined sample volume. This issue can be resolved by utilizing a new design to measure simultaneously the settling velocity and morphology with a larger imaging area and a potentially undisturbed vertical channel above it. Thus, correlations among settling velocity, morphology, and density can be obtained with combined data from the current and new systems. Finally, the collected data is stored in local drives, which would take up a large portion of the storage space. However, in this way, we can scan through the database of all the snow particle types after several deployments and improve the robustness and accuracy of the machine learning model. With such a robust and accurate model, we can launch the real-time data processing in parallel with the acquisition in the future.

*Acknowledgments.*

The authors thank the staff from St. Anthony Falls Laboratory, including Benjamin Erickson and Erik Steen, for their assistance in the system design and fabrication, and Nathaniel Bristow and Peter Hartford for their help during the field deployments. This work is supported by the National Science Foundation (Program Manager, Nicholas Anderson) under grant NSF-AGS-1822192. The authors declare no conflicts of interest.

*Data Availability Statement.*

APPENDIX

**Volume estimation for snow aggregates**

Due to the uniform geometry of graupels and small particles (close to a sphere), as well as disk, dendrites, and needles (close to a cylinder), we can estimate their volume easily using one or more of the measured parameters, i.e., major axis length, minor axis length, and equivalent diameter. However, snow aggregates can have more complex and irregular geometries, as shown in the figure A1a and c. Thus, the simple spherical assumption would lead to large uncertainties in volume estimation, especially for the ones with a large aspect ratio (i.e., major axis over minor axis). As the snow aggregates are composite of several snow particles (monomer), we apply image segmentation (watershed function in MATLAB) to obtain those monomers and treat each monomer as a spheroid. The volume of the snow aggregate is estimated to be the sum of the volumes of each monomer ($V_{\text{seg,eq}} = \sum_{i=1}^{N} \frac{1}{6} \pi d_{\text{eq},i}^3$). By comparing the volume estimated by this method to the volume estimated assuming a single sphere ($V_{\text{eq}} = \frac{1}{6} \pi d_{\text{eq}}^3$), we find a systematic overestimation of $V_{\text{eq}}$ (figure A1e). We believe that the volume estimates after segmentation are more accurate than the volume estimates assume a single sphere.

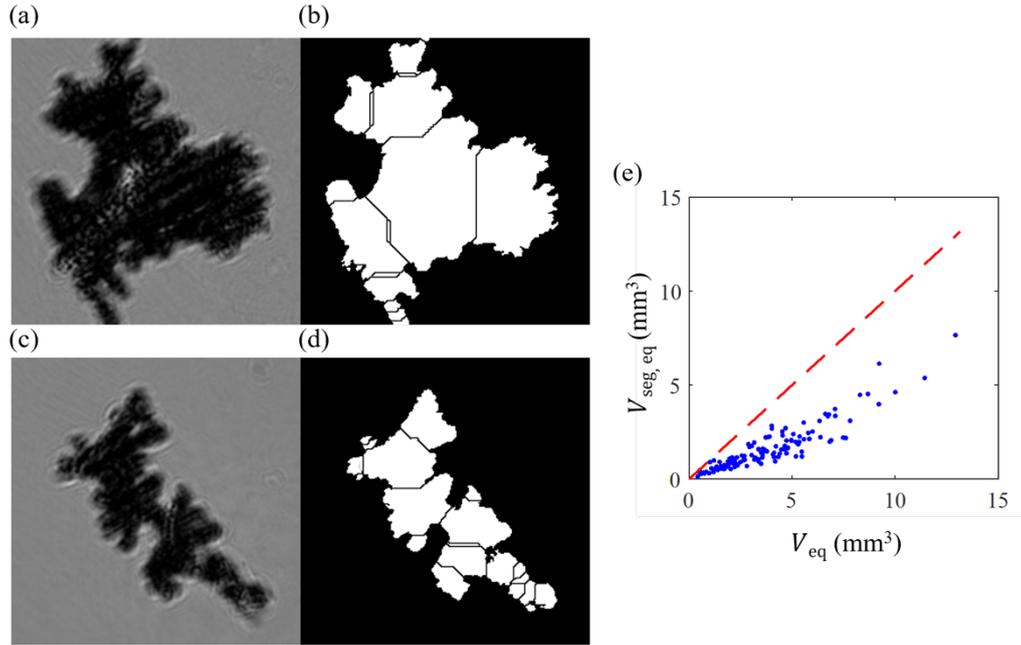

Fig. A1. (a, c) sample holograms of snow aggregates, and (b, d) the binarized image after monomer segmentation; (e) the comparison between the estimated snow aggregate volume based on equivalent diameter of the 2D projection ($V_{eq}$) and the estimated volume based on equivalent diameters of the segmented monomers ($V_{seg,eq}$).

## REFERENCES

Auer, A. H., and Veal, D. L., 1970: The dimension of ice crystals in natural clouds. *Journal of the Atmospheric Sciences*, 27(6), 919-926.

Barthazy, E., Göke, S., Schefold, R., and Högl, D., 2004: An optical array instrument for shape and fall velocity measurements of hydrometeors. *Journal of Atmospheric and Oceanic Technology*, 21(9), 1400-1416.

Beaumont, R. T., and Work, R. A., 1963: Snow sampling results from three samplers. *Hydrological Sciences Journal*, 8(4), 74-78.

Bec, J., Biferale, L., Boffetta, G., Celani, A., Cencini, M., Lanotte, A., Musacchio, S., and Toschi, F., 2006: Acceleration statistics of heavy particles in turbulence. *Journal of Fluid Mechanics*, 550, 349-358.

Cohen, J., and Rind, D., 1991: The effect of snow cover on the climate. *Journal of Climate*, 4(7), 689-706.

Clift, R., Grace, J. R., and Weber, M. E., 2005: *Bubbles, drops, and particles*. Dover Publications, Inc.

Conger, S. M., and McClung, D. M., 2009: Comparison of density cutters for snow profile observations. *Journal of Glaciology*, 55(189), 163-169.

Diwan, T., Anirudh, G., & Tembhurne, J. V. (2022). Object detection using YOLO: challenges, architectural successors, datasets and applications. *Multimedia Tools and Applications*, 1-33.

Dunnavan, E. L., Jiang, Z., Harrington, J. Y., Verlinde, J., Fitch, K., and Garrett, T. J., 2019: The shape and density evolution of snow aggregates. *Journal of the Atmospheric Sciences*, 76(12), 3919-3940.

Elder, B., Shoop, S., Feyrer, M., and Beal, S., 2019: Methods for Measuring Snow Moisture and Density. *In Cold Regions Engineering 2019* (pp. 246-253). Reston, VA: American Society of Civil Engineers.

Fitch, K. E., and Garrett, T. J., 2022: Measurement and Analysis of the Microphysical Properties of Arctic Precipitation Showing Frequent Occurrence of Riming. *Journal of Geophysical Research: Atmospheres*, 127(7), e2021JD035980.

Fugal, J. P., Shaw, R. A., Saw, E. W., and Sergeyev, A. V., 2004: Airborne digital holographic system for cloud particle measurements. *Applied Optics*, 43(32), 5987-5995.

Fujiyoshi, Y., and Wakahama, G., 1985: On snow particles comprising an aggregate. *Journal of the Atmospheric Sciences*, 42(15), 1667-1674.

Garrett, T. J., Bair, E. H., Fallgatter, C. J., Shkurko, K., Davis, R. E., and Howlett, D., 2012: The multi-angle snowflake camera. In *International Snow Science Workshop* (pp. 16-21).

Garrett, T. J., and Yuter, S. E., 2014: Observed influence of riming, temperature, and turbulence on the fallspeed of solid precipitation. *Geophysical Research Letters*, 41(18), 6515-6522.

Gaudfrin, F., Santos, E., Presley, D., and Berg, M. J., 2020: Time-resolved imaging of settling mineral dust aerosols with digital holography. *OSA Continuum*, 3(9), 2493-2500.

Guo, B., Nyman, L., Nayak, A. R., Milmore, D., McFarland, M., Twardowski, M. S., Sullivan, J. M., Yu, J., and Hong, J., 2021: Automated plankton classification from holographic imagery with deep convolutional neural networks. *Limnology and Oceanography: Methods*, 19(1), 21-36.

Harimaya, T., 1988: The relationship between graupel formation and meteorological conditions. *Journal of the Meteorological Society of Japan*. Ser. II, 66(4), 599-606.

Haussener, S., Gergely, M., Schneebeli, M., and Steinfeld, A., 2012: Determination of the macroscopic optical properties of snow based on exact morphology and direct pore-level heat transfer modeling. *Journal of Geophysical Research: Earth Surface*, 117(F3).

Haynes, W. M., Lide, D. R., & Bruno, T. J., 2016: *CRC Handbook of Chemistry and Physics*. CRC press.

Heymsfield, A., 1972: Ice crystal terminal velocities. *Journal of Atmospheric Sciences*, 29(7), 1348-1357.

Heymsfield, A. J., Bansemer, A., Schmitt, C., Twohy, C., and Poellot, M. R., 2004: Effective ice particle densities derived from aircraft data. *Journal of the Atmospheric Sciences*, 61(9), 982-1003.

Hribik, M., Vida, T., Skvarenina, J., Skvareninova, J., and Ivan, L., 2012: Hydrological effects of Norway spruce and European beech on snow cover in a mid-mountain region of the Polana mts., Slovakia. *Journal of Hydrology and Hydromechanics*, 60(4), 319-332.

Ishizaka, M., 1993: An accurate measurement of densities of snowflakes using 3-D microphotographs. *Annals of Glaciology*, 18, 92-96.

Ishizaka, M., Motoyoshi, H., Yamaguchi, S., Nakai, S., Shiina, T., and Muramoto, K. I., 2016: Relationships between snowfall density and solid hydrometeors, based on measured size and fall speed, for snowpack modeling applications. *The Cryosphere*, 10(6), 2831-2845.

Jiang, Z., Verlinde, J., Clothiaux, E. E., Aydin, K., & Schmitt, C., 2019: Shapes and fall orientations of ice particle aggregates. *Journal of the Atmospheric Sciences*, 76(7), 1903-1916.

Jocher, G., Nishimura, K., Mineeva, T., & Vilariño, R., 2020: yolov5. Code repository.

Jonas, T., Marty, C., and Magnusson, J., 2009: Estimating the snow water equivalent from snow depth measurements in the Swiss Alps. *Journal of Hydrology*, 378(1-2), 161-167.

Katz, J., & Sheng, J., 2010: Applications of holography in fluid mechanics and particle dynamics. *Annual Review of Fluid Mechanics*, 42(1), 531-555.

Kleinkort, C., Huang, G. J., Bringi, V. N., and Notaroš, B. M., 2017: Visual hull method for realistic 3D particle shape reconstruction based on high-resolution photographs of snowflakes in free fall from multiple views. *Journal of Atmospheric and Oceanic Technology*, 34(3), 679-702.

Kumar, S. S., Sun, Y., Zou, S., and Hong, J., 2016: 3D holographic observatory for long-term monitoring of complex behaviors in drosophila. *Scientific Reports*, 6(1), 1-7.

Larsen, M. L., Shaw, R. A., Kostinski, A. B., and Glienke, S., 2018: Fine-scale droplet clustering in atmospheric clouds: 3D radial distribution function from airborne digital holography. *Physical Review Letters*, 121(20), 204501.

Li, C., Lim, K., Berk, T., Abraham, A., Heisel, M., Guala, M., Coletti, F., and Hong, J., 2021: Settling and clustering of snow particles in atmospheric turbulence. *Journal of Fluid Mechanics*, 912.

Li, J., Abraham, A., Guala, M., and Hong, J., 2021: Evidence of preferential sweeping during snow settling in atmospheric turbulence. *Journal of Fluid Mechanics*, 928.

Lin, T. Y., Maire, M., Belongie, S., Hays, J., Perona, P., Ramanan, D., Dollár, P., & Zitnick, C. L., 2014: Microsoft COCO: Common Objects in Context. In *European Conference on Computer Vision* (pp. 740-755). Springer, Cham.

Magono, C., and Lee, C. W., 1966: Meteorological classification of natural snow crystals. *Journal of the Faculty of Science*, Hokkaido University. Series 7, Geophysics, 2(4), 321-335.

Marks, D., Kimball, J., Tingey, D., and Link, T., 1998: The sensitivity of snowmelt processes to climate conditions and forest cover during rain-on-snow: A case study of the 1996 Pacific Northwest flood. *Hydrological Processes*, 12(10-11), 1569-1587.

Nemes, A., Dasari, T., Hong, J., Guala, M., and Coletti, F., 2017: Snowflakes in the atmospheric surface layer: observation of particle-turbulence dynamics. *Journal of Fluid Mechanics*, 814, 592.

Ogura, T., Kageyama, I., Nasukawa, K., Miyashita, Y., Kitagawa, H., and Imada, Y., 2002: Study on a road surface sensing system for snow and ice road. *JSAE Review*, 23(3), 333-339.

Praz, C., Roulet, Y. A., and Berne, A., 2017: Solid hydrometeor classification and riming degree estimation from pictures collected with a Multi-Angle Snowflake Camera. *Atmospheric Measurement Techniques*, 10(ARTICLE), 1335-1357

Proksch, M., Löwe, H., & Schneebeli, M., 2015: Density, specific surface area, and correlation length of snow measured by high‐resolution penetrometry. *Journal of Geophysical Research: Earth Surface*, 120(2), 346-362.

Proksch, M., Rutter, N., Fierz, C., and Schneebeli, M., 2016: Intercomparison of snow density measurements: bias, precision, and vertical resolution. *The Cryosphere*, 10, 371-384.

Redmon, J., Divvala, S., Girshick, R., & Farhadi, A., 2016: You only look once: Unified, real-time object detection. In *Proceedings of the IEEE Conference on Computer Vision and Pattern Recognition* (pp. 779-788).

Rees, K. N., Singh, D. K., Pardyjak, E. R., and Garrett, T. J., 2021: Mass and density of individual frozen hydrometeors. *Atmospheric Chemistry and Physics*, 21(18), 14235-14250.

Shao, S., Mallery, K., Kumar, S. & Hong, J., 2020a: Machine learning holography of 3D particle field imaging. *Optics Express*, 28(3), 2987-2999.

Shao, S., Mallery, K. & Hong, J., 2020b: Machine learning holography for measuring 3D particle distribution. *Chemical Engineering Science*, 225, 115830.

Sihvola, A., & Tiuri, M., 1986: Snow fork for field determination of the density and wetness profiles of a snow pack. *IEEE Transactions on Geoscience and Remote Sensing*, (5), 717-721.

Singh, D. K., Donovan, S., Pardyjak, E. R., and Garrett, T. J., 2021: A differential emissivity imaging technique for measuring hydrometeor mass and type. *Atmospheric Measurement Techniques*, 14(11), 6973-6990.

Steinkogler, W., Sovilla, B., and Lehning, M., 2014: Influence of snow cover properties on avalanche dynamics. *Cold Regions Science and Technology*, 97, 121-131.

Sturm, M., Holmgren, J., König, M., and Morris, K., 1997: The thermal conductivity of seasonal snow. *Journal of Glaciology*, 43(143), 26-41.

Sturm, M., Taras, B., Liston, G. E., Derksen, C., Jonas, T., and Lea, J., 2010: Estimating snow water equivalent using snow depth data and climate classes. *Journal of Hydrometeorology*, 11(6), 1380-1394.

Tagliavini, G., McCorquodale, M., Westbrook, C., Corso, P., Krol, Q., and Holzner, M., 2021: Drag coefficient prediction of complex-shaped snow particles falling in air beyond the Stokes regime. *International Journal of Multiphase Flow*, 140, 103652.

Tagliavini, G., Khan, M. H., McCorquodale, M., Westbrook, C., and Holzner, M., 2022: Wake characteristics of complex-shaped snow particles: Comparison of numerical simulations with fixed snowflakes to time-resolved particle tracking velocimetry experiments with free-falling analogs. *Physics of Fluids*, 34(5), 055112.

Toloui, M., and Hong, J., 2015: High fidelity digital inline holographic method for 3D flow measurements. *Optics Express*, 23(21), 27159-27173.

Weast, R. C., 1981: *CRC Handbook of Chemistry and Physics 61st Edtion*. Chemical Rubber Co.

You, J., Mallery, K., Mashek, D. G., Sanders, M., Hong, J., and Hondzo, M., 2020: Microalgal swimming signatures and neutral lipids production across growth phases. *Biotechnology and Bioengineering*, 117(4), 970-980.

Zhang, T., Su, H., and Wang, K., 2017, August 8-10: Comparison of snow density measurements using different equipment. In *First Workshop on NASA SnowEx Results*, Longmont, Colorado. https://snow.nasa.gov/sites/default/files/Zhang_080917_15_Snow%20Density%20Comparison.pdf